\documentstyle[%
aps,%
prl,%
twocolumn,%
psfig%
]{revtex}

\tighten

\begin{document}

\noindent{\bf Comment on "Conductance fluctuations in meso-\\
scopic normal-metal/superconductor samples"}

\vspace{\baselineskip}

Recently, Hecker {\em et al.} \cite{Heck97} experimentally studied
magnetoconductance fluctuations in a mesoscopic Au wire connected
to a superconducting Nb contact. 
They compared the rms magnitude of these conductance fluctuations in the 
superconducting state  (rms($G_{NS}$)) to that in the normal state 
(rms($G_{N}$)) by increasing the magnetic field above the critical 
field of 2.5 T.
It was reported that rms($G_{NS}$) was about 
$2.8\pm0.4$ times larger than rms($G_{N}$),
which should confirm the theoretical predicted
enhancement factor of $2\sqrt{2}\simeq2.8$.
 
In this Comment, we show that their claim is not justified.
Although not explicitly mentioned in Ref.~\cite{Heck97},
we have to assume that the rms($G$) was calculated according to:
 $\mbox{\rm rms}(G) = \mbox{\rm rms}(R)/R^2$, 
where  rms($R$) denotes the rms magnitude of the measured 
resistance fluctuations and $R$ the total measured resistance.
The point we want to make is that the authors did not take into 
account the presence of an incoherent series resistance $R_{series}$ 
from the contacts, 
which is different when the Nb is in the superconducting or normal state.
Since the measured  rms($R$) only originates from the phase-coherent
part of the disordered conductor, with resistance $R_\varphi$, 
the correct procedure is to calculate rms($G$) according to:
$\mbox{\rm rms}(G) = \mbox{\rm rms}(R)/R_\varphi^2=
\mbox{\rm rms}(R)/(R-R_{series})^2$. 
As shown below, when we correct for the presence of this series resistance,
we find that rms($G_{NS}$) is {\bf not} significantly larger than rms($G_{N}$).

Their device consists of a narrow Au wire ($Au^w$, length $L=1.0\mu$m, width
$W=0.13\mu$m) connected at its ends to a macroscopic Nb and Au contact
($Nb^c$ or $Au^c$)
via a rectangular shaped contact ($Nb^r$ or $Au^r$, $L=0.8\mu$m, $W=1.6\mu$m).
The total resistance is the sum of these five contributions:
$R = R_{Nb}^c + R_{Nb}^r + R_{Au}^w + R_{Au}^r + R_{Au}^c$,
where $ R_{Nb}^c + R_{Nb}^r$ are zero in the superconducting state.

\begin{table}[tbh]
\caption{The measured resistance $R_{\rm NS}$ and uncorrected
conductance fluctuations rms$(G_{\rm NS})$ in the superconducting state 
at T=50mK and B=1T, and the measured resistance $R_{\rm N}$ and 
the {\em corrected} conductance fluctuations rms$(G_{\rm N})$
in the normal state at T=50mK and B=4T.}
\begin{tabular}{lcc}
& sample 1 & sample 2 \\
\tableline
$R_{\rm NS}$ ($\Omega$)            & 11.60             & 9.72             \\
$R_{\rm N}$ ($\Omega$)             & 15.87             & 14.34            \\
${\rm rms}(G_{\rm NS}) $ ($e^2/h$) & 0.16  $\pm 0.02 $ & 0.14  $\pm 0.02$ \\
${\rm rms}(G_{\rm N}) $ ($e^2/h$)  & 0.109 $\pm 0.006$ & 0.109 $\pm0.009$ \\
${\rm rms}(G_{\rm NS})/{\rm rms}(G_{\rm N})$ 
				   & 1.5   $\pm 0.2 $  & 1.3   $\pm 0.2$  
\end{tabular}
\label{tab1}
\end{table}

Since the series resistances of the Au contact 
($R_{Au}^c + R_{Au}^r \simeq 1.2 R_\Box^{Au} \simeq 1.1\Omega$) 
are small compared
to phase-coherent resistance of the Au wire ($10.5 \Omega$),
we will only correct for the series resistances of the Nb contact 
($R_{Nb}^c + R_{Nb}^r \simeq 1.2 R_\Box^{Nb} \simeq 4.8 \Omega$).
This series resistance is only present in the normal state  
and is exactly equal to the increase in resistance when the magnetic field
exceeds $B_c$ (see Fig.~1 a)). 
We note that not only the macroscopic Nb contact is regarded to be incoherent, 
but the rectangular shaped Nb contact as well.
Namely, the phase-breaking length $L_\varphi\equiv\sqrt{D \tau_\varphi}$ for Nb
is expected to be reduced compared to $L_\varphi\simeq 0.6 \mu$m for Au 
by $\sqrt{D_{Au}/D_{Nb}}\simeq 2.5$, 
which implies that the resistance fluctuations from this Nb rectangle
are strongly suppressed due to ensemble-averaging as well.

In Table \ref{tab1} we have reproduced the measured (average) resistance of 
the two studied samples in the normal state and in the superconducting state.
We did not correct rms$(G_{NS})$ \cite{Hart96}.
The rms$(G_N)$ has been corrected as described above.
As a result, the rms$(G_N)$ are a factor of
$(R_{N}/R_{NS})^2\simeq2$  larger than reported in Ref.~\cite{Heck97}
and consequently the ratio rms$(G_{NS})$/rms$(G_N)$ becomes about 1.4$\pm$0.2.
We doubt, however, that the remaining difference from 1 is significant,
since the statistical error could well be larger than 0.2 due to 
the fact that only a few large fluctuations determine rms$(G_{NS})$ 
(see Fig.~1b) and Fig.~2).

In conclusion, we have argued that the measured rms$(G_{NS})$ is not 
significantly enhanced compared to rms$(G_{N})$, 
and it remains an experimental challenge to 
observe the predicted enhancement factor of $2\sqrt{2}$.

\vspace{\baselineskip}
\noindent{S.G. den Hartog and B.J. van Wees\\
Department of Applied Physics and Materials Science Centre\\
University of Groningen\\
Nijenborgh 4\\
9747 AG Groningen, The Netherlands}

\vspace{\baselineskip}

%\pacs{PACS numbers: 73.23.-b, 73.50.Jt, 74.80.-g}


\begin{thebibliography}{99}

\bibitem{Heck97}
K. Hecker, H. Hegger, A. Altland, and K. Fiegle, Phys. Rev. Lett. {\bf 79},
1547 (1997).


\bibitem{Hart96}
The reported values for rms$(G_{NS})$ are considerably smaller than the  
rms magnitude of the sample-specific conductance fluctuations 
of about rms$(G_{NS}) \simeq 1.0 e^2/h$
observed in both a cross-shaped and a T-shaped 2-dimensional electron gas 
coupled to superconductors:
S.G. den Hartog {\em et al.}, Phys. Rev. Lett. {\bf 77}, 4954 (1996);
S.G. den Hartog {\em et al.}, {\em ibid.} {\bf 76}, 4592 (1996).
A comparison with the normal state values was not made in these experiments.

\end{thebibliography}
\end{document}